\documentclass[letter,twocolumn,amsmath,amssymb,superscriptaddress]{jpsj3}
\usepackage{txfonts,amsmath,amssymb}
\usepackage{bm}
\usepackage{braket}
\usepackage{lipsum}
\usepackage{color}

\newcommand{\beginsec}[1]{\noindent \textit{#1} ---}
\newcommand{\eqn}[1]{(\ref{eqn:#1})}

\def\runauthor{J. Watanabe, Y. Araki, K. Kobayashi, A.Ozawa, K. Nomura}

\title{
    Magnetic orderings from spin-orbit coupled electrons on kagome lattice
}

\author{
    Jin Watanabe$^1$,
    Yasufumi Araki$^2$\thanks{araki.yasufumi@jaea.go.jp},
    Koji Kobayashi$^1$\thanks{k-koji@tohoku.ac.jp},
    Akihiro Ozawa$^1$\thanks{akihiro.ozawa.s4@dc.tohoku.ac.jp},
    Kentaro Nomura$^{1,3}$\thanks{kentaro.nomura.e7@tohoku.ac.jp}
}
\inst{
    $^1$Institute for Materials Research, Tohoku University, Sendai 980-8577, Japan \\
    $^2$Advanced Science Research Center, Japan Atomic Energy Agency, Tokai 319-1195, Japan \\
    $^3$Center for Spintronics Research Network, Tohoku University, Sendai 980-8577, Japan
}

\abst{
    We investigate magnetic orderings on kagome lattice numerically from the tight-binding Hamiltonian of electrons,
    governed by the filling factor and spin-orbit coupling (SOC) of electrons.
    We find that even a simple kagome lattice model can host
    both ferromagnetic and noncollinear antiferromagnetic orderings depending on the electron filling,
    reflecting gap structures in the Dirac and flat bands characteristic to the kagome lattice.
    Kane--Mele- or Rashba-type SOC tends to stabilize noncollinear orderings,
    such as magnetic spirals and 120-degree antiferromagnetic orderings,
    due to the effective Dzyaloshinskii--Moriya interaction from SOC.
    The obtained phase structure helps qualitative understanding of magnetic orderings in various
    kagome-layered materials with Weyl or Dirac electrons.
}

\begin{document}
\maketitle

\beginsec{Introduction} Kagome lattice is one of the most common two-dimensional lattice structures appearing in layered crystals,
which hosts various characteristic features of electrons and magnetism \cite{Mielke_1991,Mielke_1992,Tasaki_1992,    Sachdev_1992,Lecheminant_1997,Tanaka_2003,Guo_2009,Balents_2010,Han_2012, Legendre_2020,Kim_2019, Kobayashi_robust_2019}.
The electronic states on kagome lattice show flat bands and gapless Dirac points.
They induce characteristic shapes of the Fermi surface
that can cause magnetism \cite{Barros_2014}.
Therefore, the magnetic ordering strongly depends on the Fermi level.
In other words,
a tuning of the electron filling may help us design magnetic orderings in kagome layered materials
\cite{Kassem_2015,Thakur_2020,Yanagi_2021,Ozawa_2021}.

Spin-orbit coupling (SOC) is also a fundamental factor in understanding magnetism.
Because of the correlation between the electron motion and the electron spin,
SOC should strongly affect magnetic orderings in connection with the electronic band structure on the kagome lattice.
In particular,
SOC breaks spin symmetry and leads to magnetic anisotropy\cite{Daalderop_1990},
which is one of the significant magnetic properties for spintronics devices
\cite{Kim_2019, Kobayashi_robust_2019}.
Therefore in the kagome lattice systems,
we expect more diverse magnetic orderings by
tuning SOCs\cite{Premasir_2019} in addition to the electron filling.

Recent theoretical and experimental studies have discovered various kagome-layered magnetic materials
having topological electronic states due to SOC.
Each species shows a unique magnetic ordering distinct from the others.
$\mathrm{Mn_3 Sn}$ shows a 120-degree noncollinear antiferromagnetic
(AFM)
ordering at room temperature,
with Weyl points in the electronic band structure
\cite{Nakatsuji_2015,Kuroda_2017,Liu_2017,Ito_2017,Higo_2018,Higo_2018_2,Park_2018}.
Despite its small net magnetization,
it shows the strong anomalous Hall effect (AHE)
due to the Berry curvature \cite{Nagaosa_2010,Xiao_2010,Zhang_2011,Chen_2014,Zhang_SO_2020}
from the Weyl points.
$\mathrm{Co_3 Sn_2 S_2}$ with the shandite structure also has Weyl points yielding the AHE,
while the Co atoms in kagome layers form an out-of-plane
(OOP) ferromagnetic (FM) ordering
\cite{Liu_2018,Wang_2018,Liu_2019,Ozawa_2019,Ikeda_2020}.
$\mathrm{Fe_3 Sn_2}$ shows an in-plane (IP) FM ordering,
in association with massive Dirac electrons and the large AHE
\cite{Ye_2018,Lin_2018, Fang_2022}.
Here alloys of $\mathrm{Fe}$ and $\mathrm{Sn}$ also form kagome bilayers of $\mathrm{Fe_3 Sn}$ with $\mathrm{Sn}$ atoms in between.
Although all of these materials commonly have kagome lattice structure,
various magnetism arise from the difference in the compositions, 
which give different electron numbers.
To explain the origins of these magnetic orderings in kagome materials,
we need to understand magnetic interactions derived from electronic properties.

\begin{figure}[tbp]
    \includegraphics[width=1\linewidth]{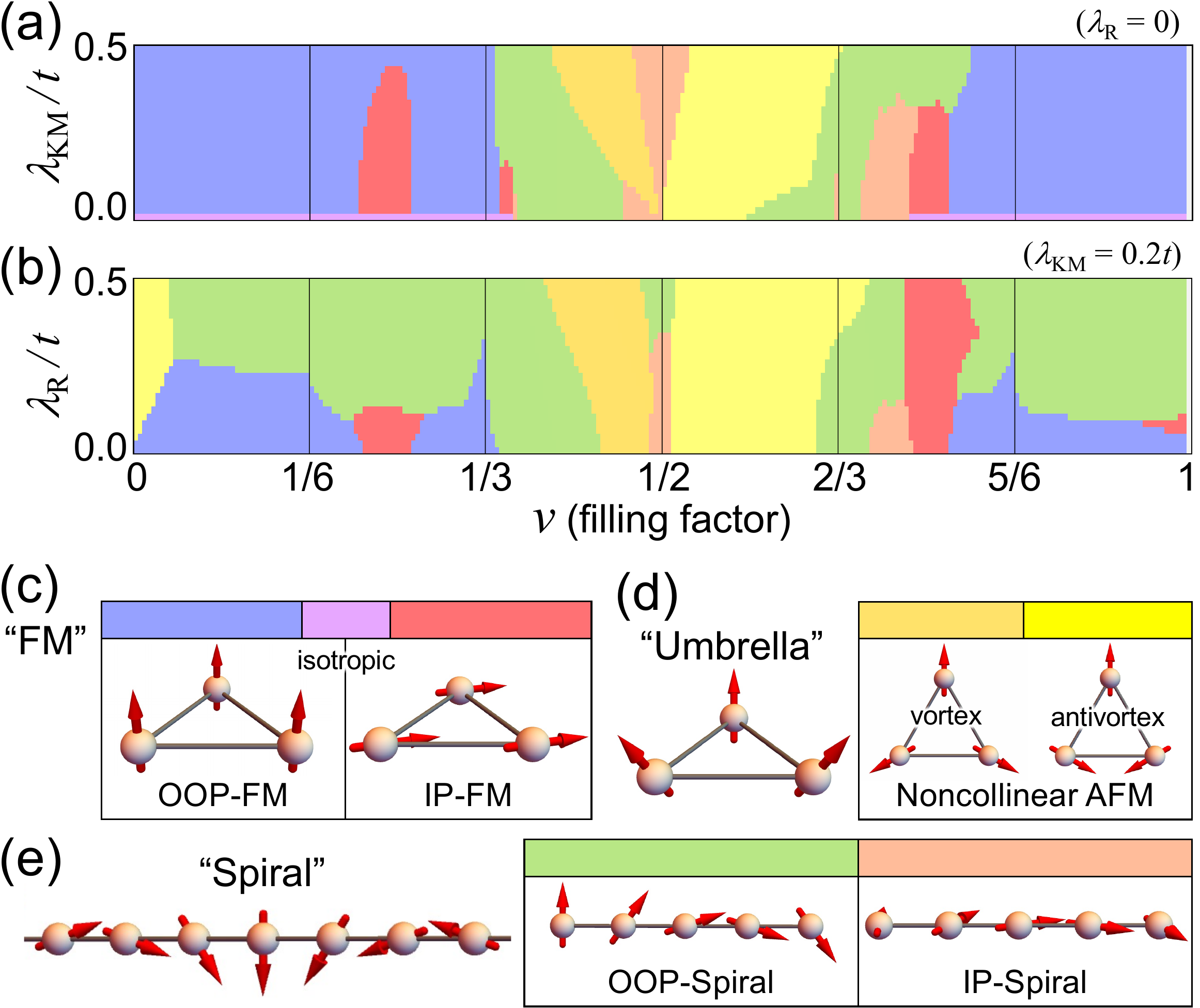}
    \caption{(Color online)
        Phase diagrams of the ground-state magnetic texture on the monolayer kagome lattice.
        We vary the filling factor $\nu$ of the electrons
        and observe (a) the dependence on the strength of the Kane--Mele-type spin-orbit coupling
        $\lambda_{\mathrm{KM}}$ with fixed $\lambda_{\mathrm{R}} =0$
        and (b) that of the Rashba-type spin-orbit coupling
        $\lambda_{\mathrm{R}}$ with fixed $\lambda_{\mathrm{KM}} =0.2t$.
        Legends with schematic pictures of the possible magnetic textures in the phase diagrams:
        (c) ferromagnetic, (d) umbrella or noncollinear antiferromagnetic, and (e) spiral orderings.
    }
    \label{fig:phase-diagram}
\end{figure}

In this article, we study the behavior of magnetic orderings on the kagome lattice from the electronic band structures.
Starting from the microscopic Hamiltonian of electrons coupled with localized magnetic moments on the kagome lattice,
we evaluate the energies of the electron systems under a variety of magnetic orderings.
Then we determine the ground-state magnetic ordering among them,
which we map into phase diagrams
by varying the number of electrons and the strengths of SOCs,
including the Kane--Mele (KM) type and the Rashba type.
The resulting phase diagrams are shown in Fig.~\ref{fig:phase-diagram}.
The phase diagrams host the OOP- and IP-FM orderings,
the vortex- and antivortex-like noncollinear AFM orderings,
and also the magnetic spirals.
The FM orderings appear away from the half filling,
whereas the noncollinear AFM orderings appear and flip their vorticity around the half filling.
Furthermore,
in the presence of the Rashba-type SOC,
the magnetic spiral ordering
\cite{Kaplan_1961,Sosnowska_1982} emerges.
To understand the origins of the magnetic orderings from the viewpoint of spin systems,
we derive an effective model for classical localized spins.
The model includes the Heisenberg interaction, magnetic anisotropy, and the Dzyaloshinskii--Moriya (DM) interactions
\cite{Dzyaloshinskii_1958,Moriya_1960,Rigol_2007}.
By focusing on the gap structure and the density of states of the electrons,
we show that the obtained phase diagrams and the effective spin model
can be understood qualitatively from the electronic band structure characteristic to the kagome lattice.

\beginsec{Model}
For our numerical calculations,
we use the kagome monolayer model that hosts both electrons and localized magnetic moments on the kagome sites \cite{Supplemental}.
The model is defined as a tight-binding Hamiltonian composed of three parts,
\begin{align} \label{eqn:H}
    H = H_{\mathrm{hop}} + H_{\mathrm{SOC}} + H_{\mathrm{exc}}.
\end{align}
Here the constituent terms represent
the electron hopping,
the effect of SOC,
and the exchange coupling between the electrons and localized magnetic moments,
respectively.
With the annihilation operator
$c_{i} = (c_{i\uparrow}, c_{i\downarrow})$
and creation operator $c_i^\dag$ of the electrons of spin-$\uparrow$ and $\downarrow$ at kagome site $i$,
the hopping term is defined by
\begin{align}
    H_{\mathrm{hop}} &= t \sum_{\langle ij \rangle} c_{i}^\dag c_{j},
\end{align}
which we restrict to the nearest neighboring sites $\langle ij \rangle$.
As is well known, this tight-binding model gives a flat band and a pair of the Dirac points.
We add to this model the effect of SOC,
\begin{align}
    H_{\mathrm{SOC}} &= i\lambda_{\mathrm{KM}} \sum_{\langle \! \langle ij \rangle \! \rangle} \nu_{ij} c_i^\dag \sigma_z c_j + i \lambda_{\mathrm{R}} \sum_{\langle ij \rangle} c_i^\dag (\boldsymbol{\sigma} \times \boldsymbol{e}_{ij})_z c_j.
\end{align}
The first term describes the KM-type SOC
\cite{KaneMele,Guo_2009}
arising from the local breaking of inversion symmetry,
which acts between the next-nearest neighboring sites $\langle \! \langle ij \rangle \! \rangle$ and is odd under inversion, $\nu_{ij} = -\nu_{ji} (= \pm 1)$.
This KM-type SOC preserves spin $\sigma_z$ and opens a gap at the Dirac points \cite{Guo_2009}.
The second term corresponds to the Rashba-type SOC occurring at surfaces or interfaces,
which acts as an effective magnetic field perpendicular to the unit vector $\boldsymbol{e}_{ij}$ between the nearest neighboring sites $\langle ij \rangle$.
This Rashba-type term breaks the $\sigma_z$ conservation and
correlates the IP spin degrees of freedom with the electron motion.
Finally, we introduce the exchange coupling,
\begin{align}
    H_{\mathrm{exc}} &= -J_H S\sum_i \boldsymbol{n}_i \cdot c_i^\dag \boldsymbol{\sigma} c_i.
\end{align}
We treat the magnetic moment on each site $i$ as a classical spin,
with its amplitude $S$ and direction $\boldsymbol{n}_i$,
and couple it to the electron spin on the same site.
 In the following calculations,
 we set $J_H S = 3.5t$,
 which makes the itinerant electron states largely spin polarized
 and splits the energies of the spin-up and down bands.
 This setting may account for a strong Hund's coupling
 arising from the high-spin states composed of localized $d$-electrons.

\begin{figure}[tbp]
    \includegraphics[width=1.0\linewidth]{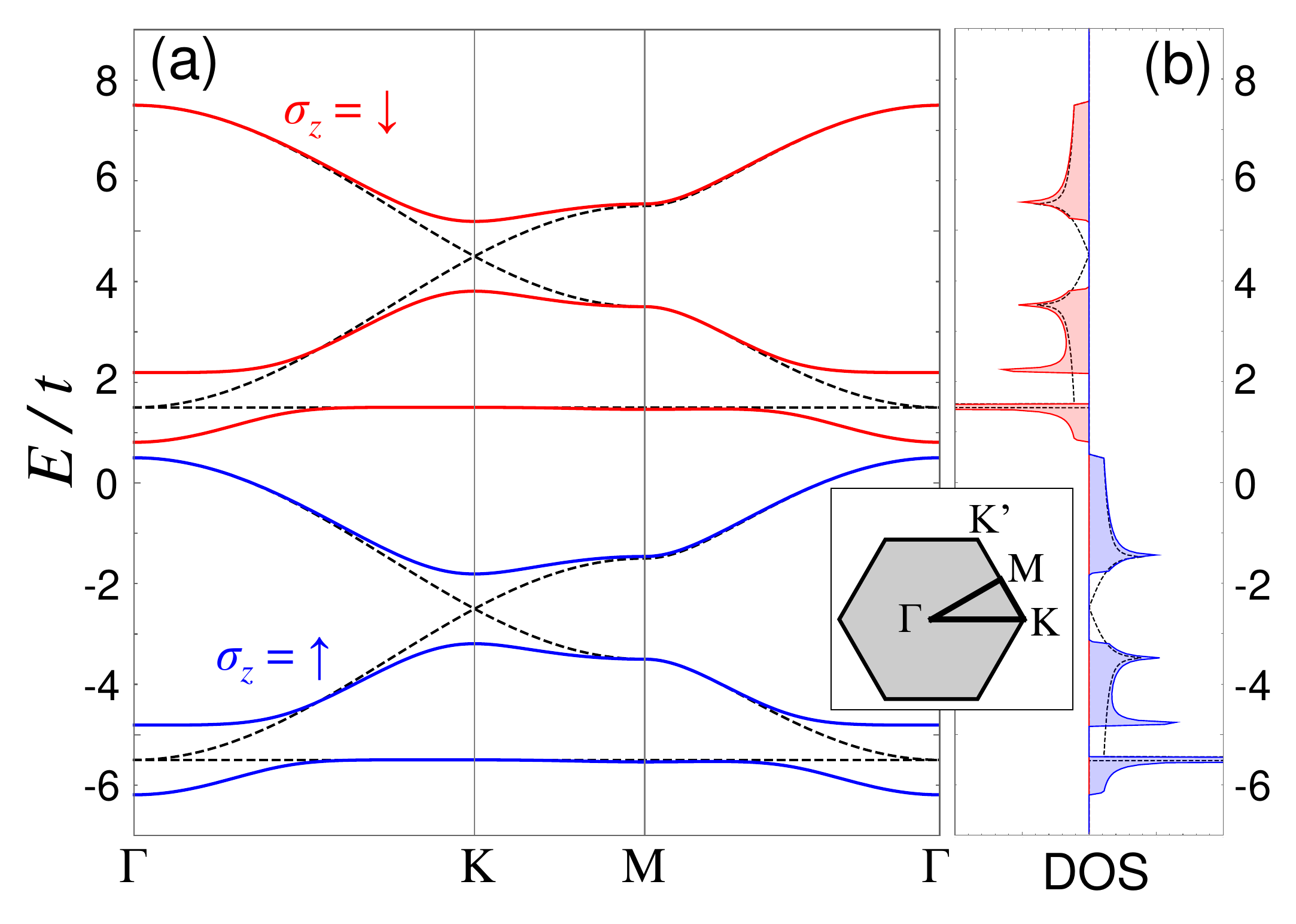}
    \caption{(Color online)
        (a) Band structure and (b) spin-resolved density of states
        under the out-of-plane ferromagnetic ordering $\boldsymbol{n}_i = \hat{\boldsymbol{z}}$
        with $J_H S = 3.5t$.
        The black dashed lines show the bands and the density of states without spin-orbit coupling,
        whereas the red and blue solid lines show those with Kane--Mele-type spin-orbit coupling $\lambda_{\mathrm{KM}} = 0.2 t$.
    }
    \label{fig:bands}
\end{figure}

Under the uniform OOP-FM ordering $\boldsymbol{n}_i = \hat{\boldsymbol{z}}$ without SOC,
the spin-up and spin-down states are energetically split,
and hence the band structure and the spin-resolved density of states are given as shown
by the black dashed lines in Fig.~\ref{fig:bands}.
The system has two flat bands showing the large density of states and
four bands forming the Dirac points at $K$ and $K'$ points.
Once we introduce the KM-type SOC $\lambda_{\mathrm{KM}}$,
it opens bandgaps between these bands,
and the flat bands become weakly dispersed,
as shown by the blue and red dashed lines in Fig.~\ref{fig:bands}.
Since the model consists of six bands,
the filling factor $\nu = 1/6$ yields complete filling of the lower-energy flat band,
and $\nu = 2/6$ yields the Fermi level at the lower-energy Dirac points,
whereas $\nu = 4/6$ and $5/6$ lead to those for the upper-energy bands.
The half filling $\nu = 3/6$ corresponds to the complete filling of all the three lower-energy bands.
We treat those filling factors as the representative filling factors,
in the discussions on our calculation results below.

\beginsec{Phase diagram}
With the tight-binding model defined above,
we calculate the total energy of the system under a given magnetic texture $\{ \boldsymbol{n}_i \}$,
by summing the eigenenergies of all the occupied electronic states.
By comparing the total energies for various magnetic textures,
we determine the ground-state magnetic texture for a given filling factor $\nu$ (at zero temperature).
As typical magnetic textures possible on a monolayer kagome lattice,
we compare three types of magnetic textures as schematically shown in Fig.~\ref{fig:phase-diagram}:
(c) the uniform FM ordering,
(d) the ``umbrella'' structure\cite{Ohgushi_2000},
and (e) the ``spiral'' structure extending periodically to one spatial direction
\cite{Supplemental}.
The umbrella structure consists of ferromagnetically aligned OOP components and noncollinearly aligned IP components,
where the IP components form either the vortex-like or the antivortex-like structure within each triangular unit cell.
It reduces to the noncollinear AFM ordering
if its opening angle $\theta$ reaches $\pi/2$
(see Supplemental Material).

By identifying the ground-state magnetic texture for every set of parameters $\nu$ and $(\lambda_{\mathrm{KM}},\lambda_{\mathrm{R}})$,
we obtain the phase diagrams
as shown in Figs.~\ref{fig:phase-diagram}(a) and \ref{fig:phase-diagram}(b).
These are the main results in this article,
where we vary $\lambda_{\mathrm{KM}}$ with $\lambda_{\mathrm{R}} =0$ fixed in panel (a),
and vary $\lambda_{\mathrm{R}}$ with $\lambda_{\mathrm{KM}} =0.2t$ fixed in panel (b).
The characteristics of the obtained phase diagrams can be described by the following three statements:
(i) The OOP-FM ordering arises for the fillings $\nu \lesssim 1/3$ and $\gtrsim 2/3$.
(ii) The noncollinear AFM ordering arises around the half filling $\nu \approx 1/2$
with either the vortex-like or the antivortex-like structure.
(iii) The SOC parameters $\lambda_{\mathrm{KM}}$ and $\lambda_{\mathrm{R}}$
both stabilize the noncollinear (AFM and spiral) orderings.

Let us explain the characteristics of the obtained phase diagrams in more detail.
First, in the absence of the SOC term [$\lambda_{\mathrm{R}} = \lambda_{\mathrm{KM}} =0$, Fig.~\ref{fig:phase-diagram}(a)],
we find an isotropic FM ordering for the fillings $\nu \lesssim 1/3$ and $\gtrsim 2/3$,
and the noncollinear AFM ordering for $\nu \approx 1/2$.
Once we switch on the KM-type SOC $\lambda_{\mathrm{KM}}$,
the FM ordering points to the OOP direction in most regions,
while the noncollinear AFM ordering becomes stabilized and expands around $\nu \approx 1/2$.
When we increase the strength of the Rashba-type SOC $\lambda_{\mathrm{R}}$
[Fig.~\ref{fig:phase-diagram}(b)],
the noncollinear AFM regions are slightly extended,
while the OOP-FM ordering
($\nu \lesssim 1/3$ and $\gtrsim 5/6$)
tends to turn into the OOP-spiral structure in most regions
once $\lambda_{\mathrm{R}}$ surpasses $\lambda_{\mathrm{KM}}$.
In the rest of this article,
we quantify those characteristics in terms of the classical spin Hamiltonian and discuss the origins of these magnetic orderings
based on the electronic band structure.

\begin{figure}[tbp]
    \includegraphics[width=\linewidth]{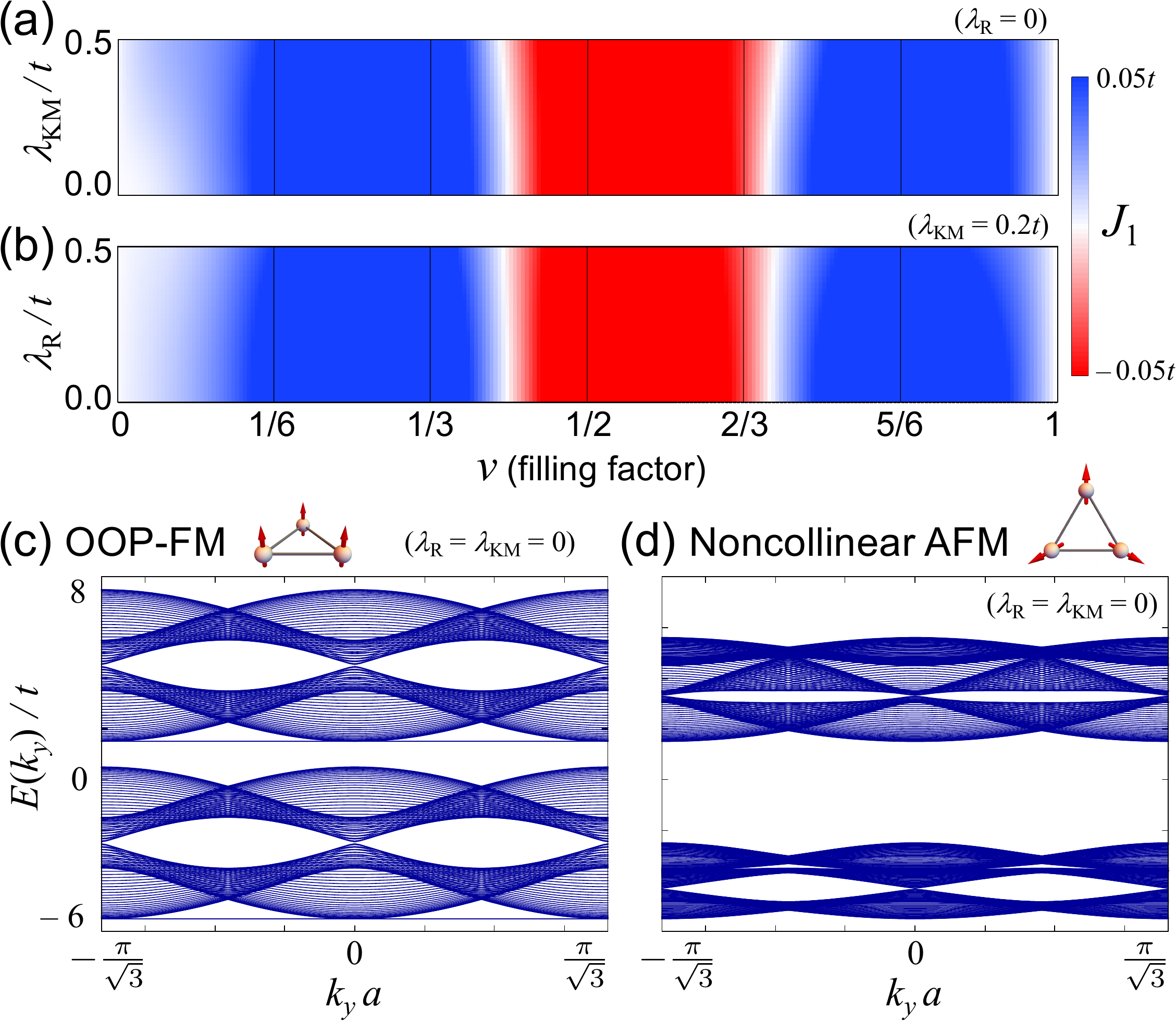}
    \caption{(Color online)
        Upper panels: The color maps of the numerically estimated effective spin-spin interaction $J_1$ between nearest neighboring sites,
        (a) in $\nu$--$\lambda_{\mathrm{KM}}$ plane with $\lambda_{\mathrm{R}} =0$ fixed
        and (b) in $\nu$--$\lambda_{\mathrm{R}}$ plane with $\lambda_{\mathrm{KM}} =0.2t$ fixed.
        Lower panels: The band structures calculated without spin-orbit coupling, under (c) the out-of-plane ferromagnetic ordering and (d) the noncollinear antiferromagnetic ordering.
    }
    \label{fig:J1}
\end{figure}

\beginsec{Ferromagnetism vs antiferromagnetism}
First we focus on the FM and noncollinear AFM ground states of our model Eq.~\eqn{H}.
These ground states depend on the filling factor $\nu$.
In order to verify the tendency of the spin system toward the FM or AFM orderings,
we estimate the strengths of the effective spin-spin interactions
by fitting the total energy calculated above to the classical spin Hamiltonian \cite{Supplemental_2}:
$-J_1 \sum_{\langle ij \rangle} \boldsymbol{n}_i \cdot \boldsymbol{n}_j$
between nearest neighboring sites
and
$-J_2 \sum_{\langle\!\langle ij \rangle\!\rangle} \boldsymbol{n}_i \cdot \boldsymbol{n}_j$
between next-nearest neighboring sites.
The estimated $J_1$ as functions of $\nu$
for varying $\lambda_{\rm KM}$ and $\lambda_{\rm R}$ are shown in Figs.~\ref{fig:J1}(a) and \ref{fig:J1}(b), respectively.
The sign-changing behavior of $J_1$ depending on the filling factor $\nu$
clearly explains the emergence of FM and AFM orderings seen in the phase diagrams [Figs.~\ref{fig:phase-diagram}(a) and \ref{fig:phase-diagram}(b)].
On the other hand, $J_1$ is almost independent of the strength of SOCs,
though it is slightly increasing with $\lambda_{\rm KM}$ [Fig.~\ref{fig:J1}(a)] and slightly decreasing with $\lambda_{\rm R}$ [Fig.~\ref{fig:J1}(b)].
We note that the magnitude of $J_1$ is about one order larger than $J_2$ \cite{Supplemental_2}.
Thus we can understand that the magnetic orderings are governed by $J_1$.
Origins of the FM and AFM orderings can be qualitatively understood
from the electronic band structure.
We compare the band structures under the OOP-FM and the noncollinear AFM orderings without SOC
in Figs.~\ref{fig:J1}(c) and \ref{fig:J1}(d),
respectively.
Due to the strong exchange interaction,
two spin states under the OOP-FM ordering are largely split in energy,
showing a flat band and Dirac points in each spin state as displayed in Fig.~\ref{fig:J1}(c).
The noncollinear AFM ordering hybridizes the spin-up and spin-down states and leads to a level repulsion,
which opens a large bandgap between the lower three bands and upper three bands as shown in Fig.~\ref{fig:J1}(d).
From those behaviors of the bands,
we can qualitatively understand how the ground-state magnetic texture depends on the filling factor $\nu$.
The filling of electrons in the low-energy flat band in the FM ordering lowers the total energy
in comparison with the AFM state.
Therefore we can understand that the FM ordering is energetically favored
in the middle of the upper or lower energy bands ($\nu \lesssim 1/3$ or $\gtrsim 2/3$).
On the other hand, the noncollinear AFM ordering favored around the half filling $(\nu \approx 1/2)$
can be traced back to the large bandgap emerging at zero energy.
The vortex-like and antivortex-like orderings are energetically degenerate in the absence of SOC.
The splitting of their degeneracy shall be discussed later in connection with the DM interaction.

\begin{figure}[tbp]
    \includegraphics[width=\linewidth]{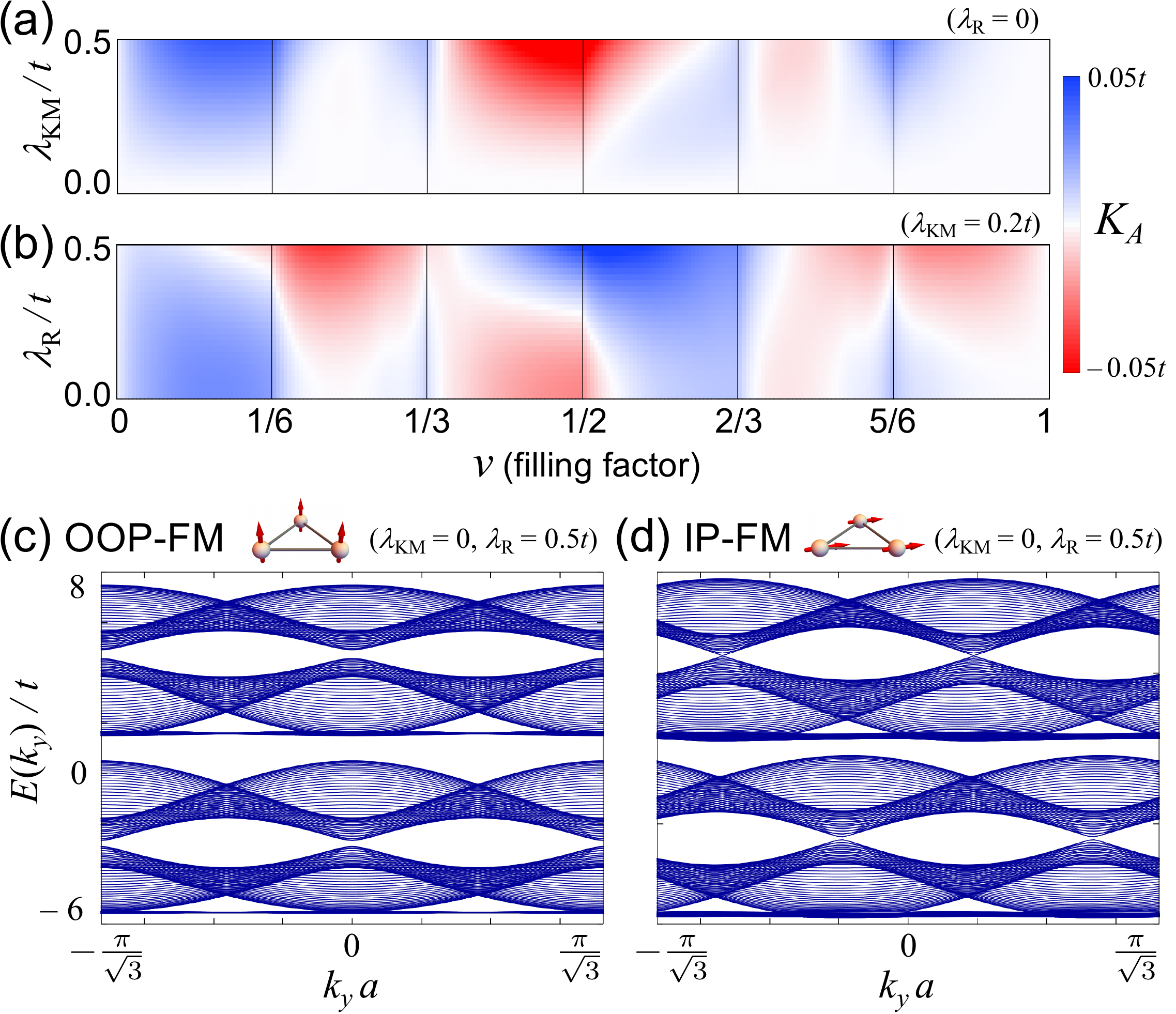}
    \caption{(Color online)
        Upper panels: The numerically obtained behavior of the magnetic anisotropy $K_A$,
        (a) with $\lambda_{\mathrm{KM}}$ varied and $\lambda_{\mathrm{R}} =0$ fixed,
        and (b) with $\lambda_{\mathrm{R}}$ varied and $\lambda_{\mathrm{KM}} =0.2t$ fixed.
        Lower panels: The band structures calculated with $\lambda_{\mathrm{R}} =0.5t$ and $\lambda_{\mathrm{KM}}=0$
        under (c) the out-of-plane ferromagnetic ordering
        and (d) the in-plane ferromagnetic ordering.
    }
    \label{fig:K}
\end{figure}

\beginsec{Magnetic anisotropy}
The SOC term $H_{\mathrm{SOC}}$ correlates the spin degrees of freedom with the in-plane motion of electrons,
which gives rise to the magnetic anisotropy.
The behavior of the magnetic anisotropy $-K_A \sum_i (n_i^z)^2$, estimated from the total energy of the system \cite{Supplemental_2},
is shown in Figs.~\ref{fig:K}(a) and \ref{fig:K}(b).
By raising the KM-type SOC $\lambda_{\mathrm{KM}}$,
we find that $K_A$ gets positively enhanced in most of the FM region ($\nu \lesssim 1/3$ and $\gtrsim 2/3$).
The enhancement of $K_A$ accounts for our finding
that the OOP-FM ordering is rather favored in the phase diagram [Fig.~\ref{fig:phase-diagram}(a)].
That can again be understood from the band structure;
as we have mentioned in Fig.~\ref{fig:bands},
$\lambda_{\mathrm{KM}}$ opens gaps above the flat bands and at the Dirac points.
Once we introduce the FM ordering with the coupling $J_H$ stronger than $\lambda_{\mathrm{KM}}$,
the OOP-FM ordering keeps the SOC gap and splits the spin-up and down bands,
whereas the IP-FM ordering closes the SOC gap [see Figs.~\ref{fig:K}(c) and \ref{fig:K}(d)].
Therefore, we can understand that the OOP-FM ordering is preferred around the flat bands
$(\nu \approx 1/6,2/3)$
and the Dirac points
$(\nu \approx 1/3,5/6)$.

The Rashba-type SOC $\lambda_{\mathrm{R}}$ also affects the magnetic anisotropy,
as shown in Fig.~\ref{fig:K}(b).
By raising the magnitude of $\lambda_{\mathrm{R}}$ in the FM region ($\nu \lesssim 1/3$ and $\gtrsim 2/3$),
we find that $K_A$ tends to change its sign from positive to negative,
which means that the easy-axis anisotropy from $\lambda_{\mathrm{KM}}$ gets suppressed and turns into the easy-plane anisotropy.
The reduction of $K_A$ accounts for the suppression of the OOP-FM ordering at large $\lambda_{\mathrm{R}}$ in the phase diagram [Fig.~\ref{fig:phase-diagram}(b)].
Such behavior of $K_A$ can be qualitatively understood from the band structure with a finite $\lambda_{\mathrm{R}}$.
By comparing the band structures under the IP-and OOP-FM orderings,
as shown in Figs.~\ref{fig:K}(c) and \ref{fig:K}(d),
we find that the bandwidth under the IP-FM ordering is larger than that under the OOP-FM ordering.
In particular, the flat bands are energetically pushed down in the presence of the IP-FM ordering.
This is why $K_A$ is reduced, and the OOP-FM ordering gets suppressed by $\lambda_{\mathrm{R}}$ for $\nu \lesssim 1/3$ and $\gtrsim 2/3$.

\begin{figure}[tbp]
    \includegraphics[width=\linewidth]{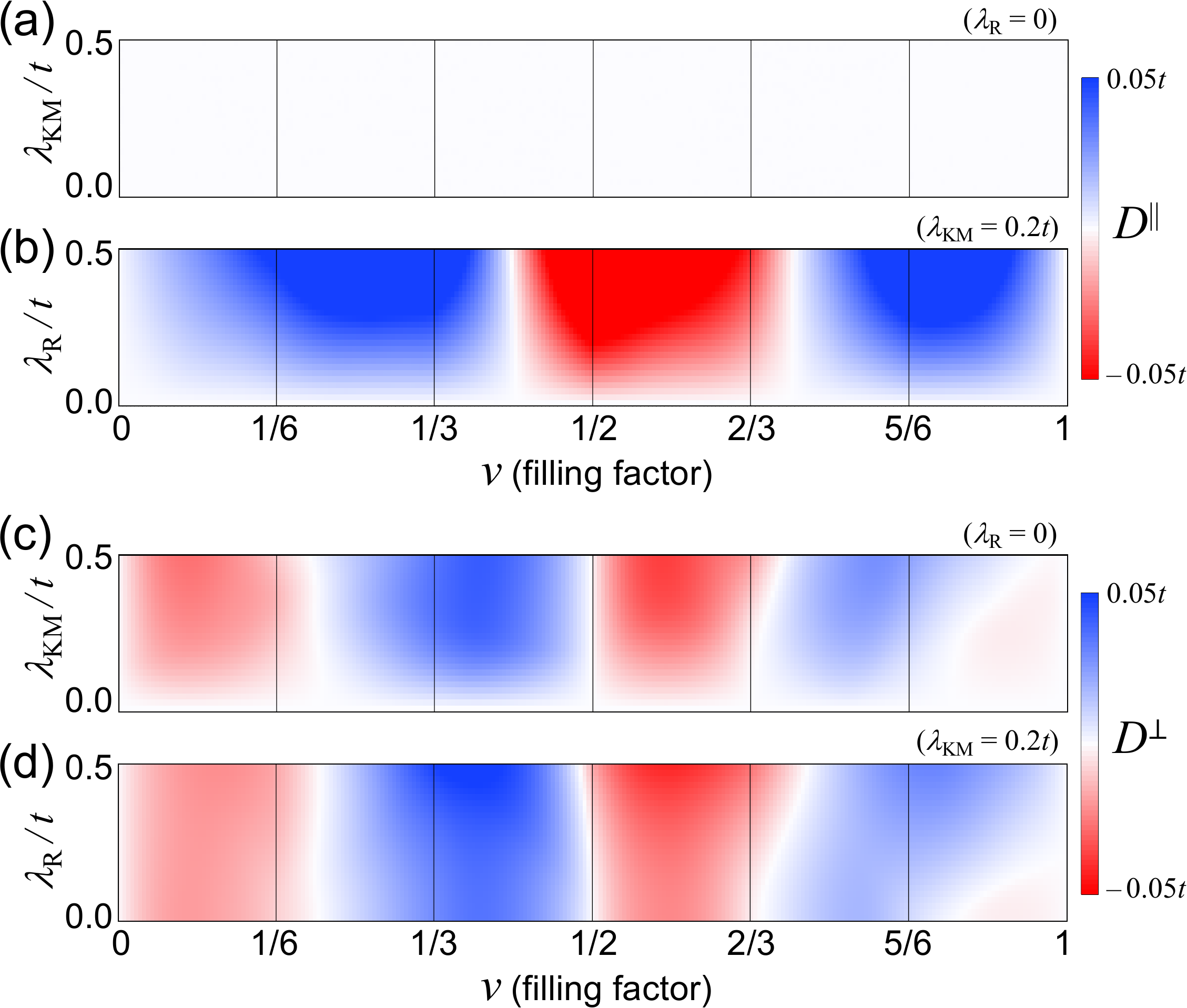}
    \caption{(Color online)
        Upper panels: The estimated strength of the in-plane component of DM interactions $D^{\parallel}$,
        (a) with $\lambda_{\mathrm{KM}}$ varied and $\lambda_{\mathrm{R}} =0$ fixed (here $D^{\parallel}$ vanishes completely),
        and (b) with $\lambda_{\mathrm{R}}$ varied and $\lambda_{\mathrm{KM}} =0.2t$ fixed.
        Upper panels: The estimated strength of the out-of-plane component of DM interactions $D^{\perp}$,
        (c) with $\lambda_{\mathrm{KM}}$ varied and $\lambda_{\mathrm{R}} =0$ fixed,
        and (d) with $\lambda_{\mathrm{R}}$ varied and $\lambda_{\mathrm{KM}} =0.2t$ fixed.
    }
    \label{fig:D}
\end{figure}

\beginsec{Noncollinear orderings and Dzyaloshinskii--Moriya interaction}
We have found from the phase diagram that the SOC term tends to stabilize the noncollinear orderings;
while $\lambda_{\mathrm{KM}}$ enhances the noncollinear AFM ordering around the half filling,
$\lambda_{\mathrm{R}}$ leads to the evolution of the spiral structure
stemming from the FM state.
In order to quantify those effects of SOC,
we estimate the strengths of the DM interaction for the localized spin moments.
We here decompose the DM interaction into two components:
the IP component
$\sum_{\langle ij \rangle} \boldsymbol{D}^{\parallel}_{ij}\cdot (\boldsymbol{n}_i \times \boldsymbol{n}_j )$
with
$\boldsymbol{D}^{\parallel}_{ij} \perp \hat{\boldsymbol{z}}$
between nearest neighboring sites $\langle ij \rangle$,
which is related to the breaking of OOP inversion symmetry in connection to $\lambda_{\mathrm{R}}$,
and the OOP component
$\sum_{\langle\!\langle ij \rangle\!\rangle} \boldsymbol{D}^{\perp}_{ij}\cdot (\boldsymbol{n}_i \times \boldsymbol{n}_j )$
with
$\boldsymbol{D}^{\perp}_{ij} \parallel \hat{\boldsymbol{z}}$
between next-nearest neighboring sites $\langle\!\langle ij \rangle\!\rangle$,
which is related to the local breaking of IP inversion symmetry in connection to $\lambda_{\mathrm{KM}}$.
The directions of the DM vectors $\boldsymbol{D}^{\parallel,\perp}_{ij}$ on each link $ij$ are determined by the Moriya's rules \cite{Moriya_1960} based on the breaking pattern of inversion symmetry,
as specified in the Supplemental Material
\cite{Supplemental_2}.
By fitting these forms of the DM interactions to the total energy of the electron system calculated above,
we estimate the values of those DM interactions $D^{\parallel,\perp}$.
The dependences of $D^{\parallel,\perp}$ on the parameters $\nu$ and $(\lambda_{\mathrm{KM}},\lambda_{\mathrm{R}})$ are shown in Fig.~\ref{fig:D}.

For the IP component $D^{\parallel}$,
we find that the Rashba-type SOC $\lambda_{\mathrm{R}}$ is essential.
As shown in Fig.~\ref{fig:D}(a),
$D^{\parallel}$ completely vanishes as long as $\lambda_{\mathrm{R}}=0$.
The magnitude of $D^{\parallel}$ rises linearly with $\lambda_{\mathrm{R}}$ in both the FM and AFM regimes
[see Fig.~S5(a) in Supplemental Material].
The emergence of $D^{\parallel}$ describes the magnetic spiral state evolving with $\lambda_{\mathrm{R}}$,
which turns from the FM ordering with $\nu \lesssim 1/3$ and $\gtrsim 2/3$ in the phase diagram [Fig.~\ref{fig:phase-diagram}(b)].
The wavelength of the spiral tends to become shorter under larger $\lambda_{\mathrm{R}}$.
In contrast, to the OOP component $D^{\perp}$,
we find that the KM-type SOC $\lambda_{\mathrm{KM}}$ gives the dominant contribution.
The estimated $D^{\perp}$ rises proportionally with $\lambda_{\mathrm{KM}}$
[see Fig.~S5(b) in Supplemental Material]
and thus stabilizes the noncollinear IP texture in the AFM regime $(\nu \approx 1/2)$ for large $\lambda_{\mathrm{KM}}$.
While the positive $D^{\perp}$ prefers the vortex-like AFM ordering,
the negative $D^{\perp}$ prefers the antivortex-like ordering.
Thus, the sign-changing behavior of $D^{\perp}$ consistently describes those two AFM orderings seen around $\nu=1/2$ in the phase diagrams.

\beginsec{Conclusion}
We studied the FM, noncollinear AFM, and magnetic spiral orderings,
from a tight-binding model with SOC terms on the monolayer kagome lattice.
These magnetic orderings are greatly governed by the tuning of the electron filling.
The Kane--Mele- and Rashba-type SOCs also play important roles in stabilizing
the noncollinear AFM and spiral orderings, respectively.
We estimated the effective DM interactions among the localized spins as the origins of these magnetic orderings.

\acknowledgments
The authors would like to thank
K.~Fujiwara,
Y.~Kato,
Y.~Motome,
K.~Nakazawa, and
A.~Tsukazaki,
for valuable discussions.
This work was supported by
JST CREST, Grant No.~JPMJCR18T2
and by
JSPS KAKENHI, Grant Nos.~
JP19K14607 
and
JP20H01830.
Y.~A.~is supported by JSPS, the Leading Initiative for Excellent Young Researchers (LEADER).
A.~O.~is supported by
GP-Spin at Tohoku university 
and by
JST SPRING, Grant No.~JPMJSP2114.

\end{document}


\title{
    Supplemental Material: Magnetic orderings from spin-orbit coupled electrons on kagome lattice
}

\newcommand{\authJW}{Jin Watanabe$^1$}
\newcommand{\authYA}{Yasufumi Araki$^2$\thanks{araki.yasufumi@jaea.go.jp}}
\newcommand{\authKK}{Koji Kobayashi$^1$\thanks{k-koji@tohoku.ac.jp}}
\newcommand{\authAO}{Akihiro Ozawa$^1$}
\newcommand{\authKN}{Kentaro Nomura$^{1,3}$\thanks{kentaro.nomura.e7@tohoku.ac.jp}}
\newcommand{\instIM}{$^1$Institute for Materials Research, Tohoku University, Sendai 980-8577, Japan}
\newcommand{\instJA}{$^2$Advanced Science Research Center, Japan Atomic Energy Agency, Tokai 319-1195, Japan}
\newcommand{\instCS}{$^3$Center for Spintronics Research Network, Tohoku University, Sendai 980-8577, Japan}

\ifjpsj
 \author{
    \authJW,
    \authYA,
    \authKK,
    \authAO, and
    \authKN
 }
 \inst{
    \instIM \\
    \instJA \\
    \instCS
 }
\else
 \author{\authJW}
 \author{\authYA}
 \author{\authKK}
 \author{\authAO}
 \author{\authKN}
 \affiliation{\instIM}
 \affiliation{\instJA}
 \affiliation{\instCS}
\fi

\maketitle

\section{Detailed structure of tight-binding model}
\begin{figure}[bp]
    \centering
        \includegraphics[width=0.7\linewidth]{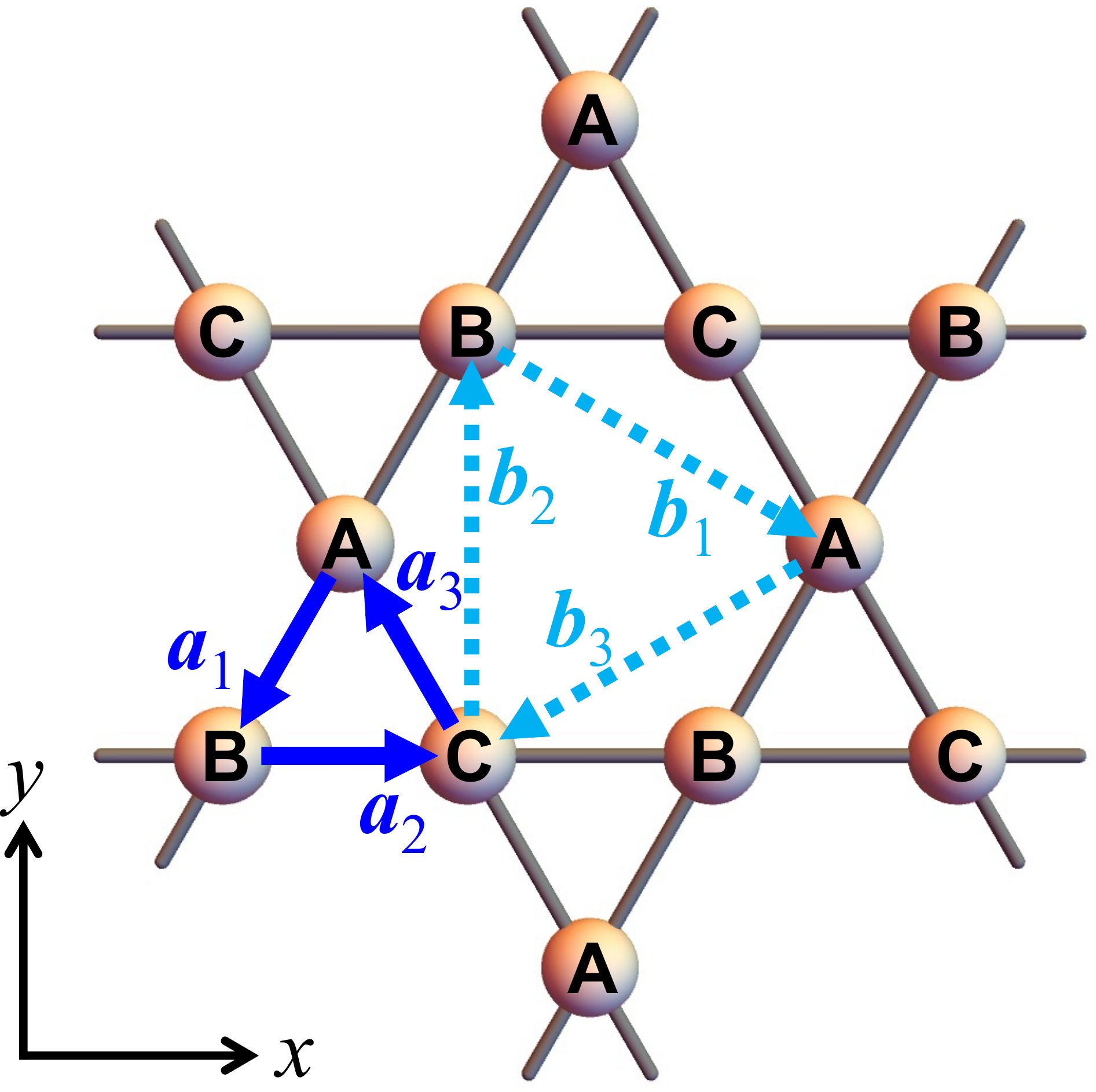}        
    \caption{
        (Color online)
        Schematic picture of the kagome lattice used for our tight-binding model.
        The vectors $\bm{a}_{1,2,3}$ denote the hoppings between nearest neighboring sites,
        and $\bm{b}_{1,2,3}$ denote the hoppings between next-nearest neighboring sites.
    }
    \label{fig:Fig_S1}
\end{figure}

In this section, we give a detailed explanation on the tight-binding model of kagome lattice used in our calculations.
The unit cell of the kagome lattice consists of three sites,
which we denote as A, B, and C (see Fig.~\ref{fig:Fig_S1}).
By denoting the distance between nearest neighboring sites as $a$,
the hoppings between nearest neighbors are characterized by the vectors
\begin{align}
    \bm{a}_1 = \left(-\frac{a}{2}, -\frac{\sqrt{3}a}{2}, 0\right), \ 
    \bm{a}_2 = (a,0,0), \ 
    \bm{a}_3 = \left(-\frac{a}{2}, \frac{\sqrt{3}a}{2}, 0\right),
\end{align}
and for next-nearest neighbors
\begin{align}
    \bm{b}_1 = \bm{a}_2-\bm{a}_3, \ 
    \bm{b}_2 = \bm{a}_3-\bm{a}_1, \ 
    \bm{b}_3 = \bm{a}_1-\bm{a}_2.
\end{align}

We define the tight-binding model on this kagome lattice,
which consists of the nearest-neighbor hopping term $H_{\mathrm{hop}}$,
the spin-orbit coupling (SOC) term $H_{\mathrm{SOC}}$,
and the exchange coupling term $H_{\mathrm{exc}}$, as explained in the main text.
In the SOC term, we introduce the Kane--Mele (KM) SOC as the imaginary hopping between next-nearest neighboring sites that is odd under spatial inversion.
This hopping is accompanied with the sign factor $\nu_{ij}$,
which takes the value $+1$ if $(i,j) = (\mathrm{A},\mathrm{B}),(\mathrm{B},\mathrm{C}),(\mathrm{C},\mathrm{A})$,
and $-1$ otherwise.
The Rashba SOC is defined between nearest neighbors with the vector $\bm{e}_{ij}$,
which is the unit vector pointing from the site $j$ to $i$.

With the above definition, $H_{\mathrm{hop}}$ and $H_{\mathrm{SOC}}$ becomes invariant under the lattice translation by a unit cell,
and hence they become diagonal in momentum space.
With the Fourier-transformed electron operator for each sublattice and spin component,
\begin{align}
    C_{\bm{k}} =
    \left(
        c_{A\uparrow,\bm{k}},\ c_{B\uparrow,\bm{k}},\ c_{C\uparrow,\bm{k}},\ \ 
        c_{A\downarrow,\bm{k}},\ c_{B\downarrow,\bm{k}},\ c_{C\downarrow,\bm{k}}
    \right),
\end{align}
we can write $H_{\mathrm{hop}}$ and $H_{\mathrm{SOC}}$ in $6\times 6$ matrix forms,
\begin{align}
    H_{\mathrm{hop}} &= \sum_{\bm{k}} C_{\bm{k}}^\dag
    \begin{pmatrix}
        \mathcal{H}_0(\bm{k}) & 0 \\
        0 & \mathcal{H}_0(\bm{k})
    \end{pmatrix}
    C_{\bm{k}}, \\
    \mathcal{H}_{0}(\bm{k}) &= 2t
    \begin{pmatrix}
        0 & \cos(\bm{k}\cdot\bm{a}_1) & \cos(\bm{k}\cdot\bm{a}_3) \\
        \cos(\bm{k}\cdot\bm{a}_1) & 0 & \cos(\bm{k}\cdot\bm{a}_2) \\
        \cos(\bm{k}\cdot\bm{a}_3) & \cos(\bm{k}\cdot\bm{a}_2) & 0  
    \end{pmatrix},
\end{align}
and
\begin{align}
    & H_{\mathrm{SOC}} = \sum_{\bm{k}} C_{\bm{k}}^\dag
    \begin{pmatrix}
        2\lambda_{\mathrm{KM}} \Gamma_{\mathrm{KM}}(\bm{k}) & -2\lambda_{\mathrm{R}} \Gamma_{\mathrm{R}} (\bm{k}) \\
        -2\lambda_{\mathrm{R}} \Gamma_{\mathrm{R}}^\dag (\bm{k}) & -2\lambda_{\mathrm{KM}} \Gamma_{\mathrm{KM}}(\bm{k})
    \end{pmatrix}
    C_{\bm{k}}, \\
    & \Gamma_{\mathrm{KM}}(\bm{k}) = \\
    & \quad \begin{pmatrix}
        0 & i\cos(\bm{k}\cdot \bm{b}_1) & -i\cos(\bm{k}\cdot \bm{b}_3) \\
        -i\cos(\bm{k}\cdot \bm{b}_1) & 0 & i\cos(\bm{k}\cdot \bm{b}_2) \\
        i\cos(\bm{k}\cdot \bm{b}_3) & -i\cos(\bm{k}\cdot \bm{b}_2) & 0
    \end{pmatrix}, \nonumber \\
    & \Gamma_{\mathrm{R}}(\bm{k}) = \\
    & \quad \begin{pmatrix}
        0 & e^{i\frac{\pi}{6}}\sin(\bm{k}\cdot\bm{a}_1) & -e^{-i\frac{\pi}{6}}\sin(\bm{k}\cdot\bm{a}_3) \\
        e^{i\frac{\pi}{6}}\sin(\bm{k}\cdot\bm{a}_1) & 0 & -i\sin(\bm{k}\cdot\bm{a}_2) \\
        -e^{-i\frac{\pi}{6}}\sin(\bm{k}\cdot\bm{a}_3) & -i\sin(\bm{k}\cdot\bm{a}_2) & 0  
    \end{pmatrix}. \nonumber
\end{align}

\section{Magnetic textures}
In this section, we give explanation about the magnetic textures investigated in our calculations in more detail.
As mentioned in the main text, we take the ``spiral'' and ``umbrella'' structures as the typical nonuniform magnetic textures that can be realized in our model.

\begin{figure}[bp]
    \centering
        \includegraphics[width=1.0\linewidth]{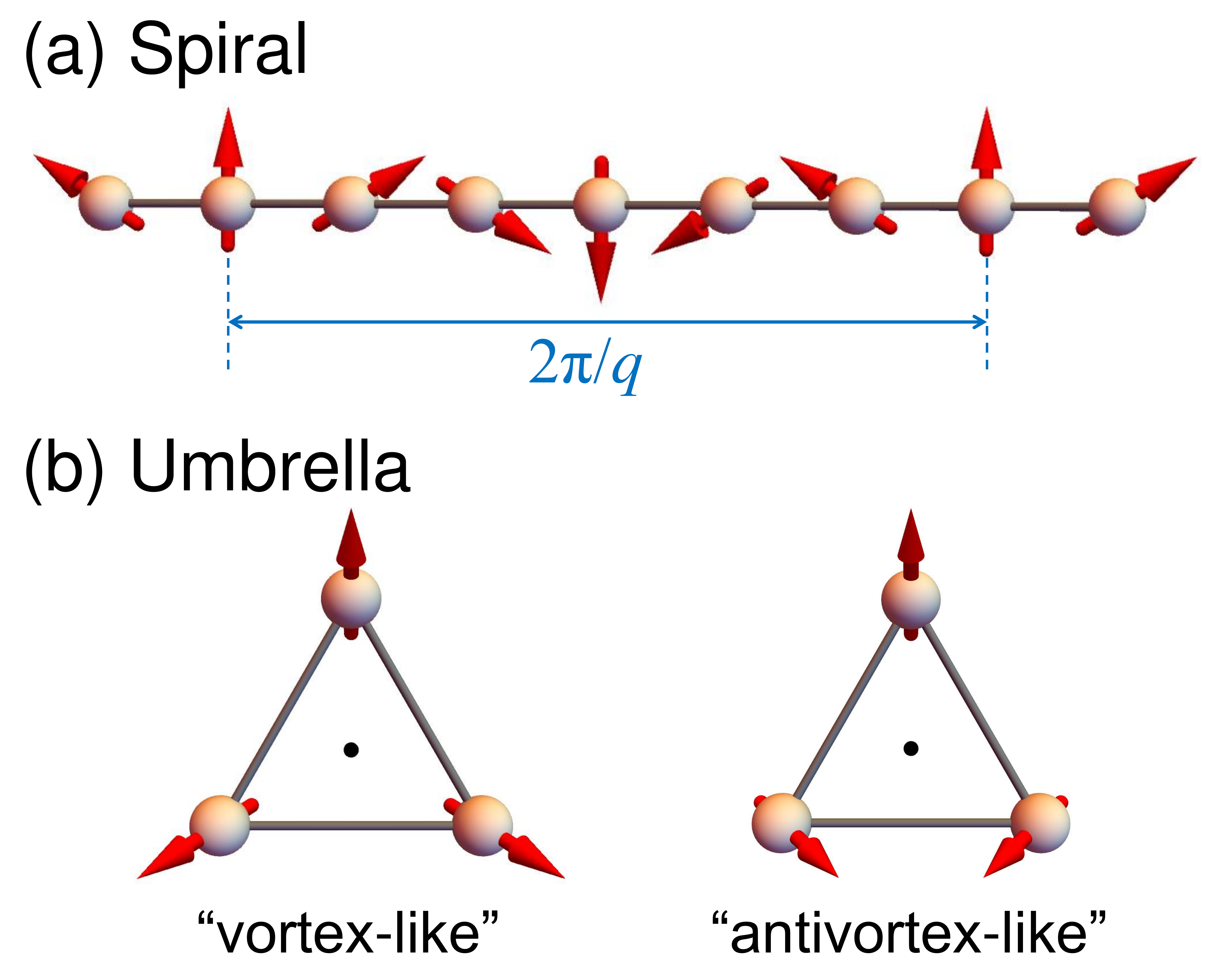}        
    \caption{
        (Color online)
        Schematic pictures of the magnetic textures investigated in our calculations: (a) the spiral structure parametrized by the wave vector $\bm{q}$,
        and (b) the ``vortex-like'' and ``antivortex-like'' umbrella structures.
    }
    \label{fig:Fig_S2}
\end{figure}

The spiral structure is the magnetic texture periodically varying in one spatial direction,
with all the spins residing in a certain plane,
as shown in Fig.~\ref{fig:Fig_S2}(a).
Its spatial variation is written as
\begin{align}
    \bm{n}(\bm{r}) = \bm{e}_1 \cos(\bm{q}\cdot\bm{r}) + \bm{e}_2 \sin(\bm{q}\cdot\bm{r}),
\end{align}
where the $\bm{q}$-vector characterizes the spatial periodicity,
and the unit vectors $\bm{e}_{1,2} \ (\bm{e}_1\cdot\bm{e}_2 =0)$ define the spiraling plane of the spins.
In our calculations,
we fix the $\bm{q}$-vector to $x$-direction,
\begin{align}
    \bm{q} = \left( 2\pi m/L_x ,0,0 \right), \quad (m \in \mathbb{Z})
\end{align}
with $L_x$ the size of the system in $x$-direction,
and take the spiraling plane spanned by $\bm{e}_{1,2}$ to $xy$-, $yz$-, or $xz$-plane.

The umbrella structure is characterized by the directions of three spins $(\bm{n}_A,\bm{n}_B,\bm{n}_C)$ within each unit cell.
Their out-of-plane (OOP) components $(n_A^z,n_B^z,n_C^z)$ are identical,
while their in-plane (IP) components $(\bm{n}_A^\parallel, \bm{n}_B^\parallel, \bm{n}_C^\parallel)$
form $120$ degrees to one another.
In particular, we consider two types of structures shown in Fig.~\ref{fig:Fig_S2}(b),
which we call the ``vortex-like'' and ``antivortex-like'' structures.
Considering the counterclockwise path $\mathrm{A \rightarrow B \rightarrow C}$ within the unit cell,
the vortex-like structure consists of the in-plane components $(\bm{n}_A^\parallel, \bm{n}_B^\parallel, \bm{n}_C^\parallel)$ aligned counterclockwise,
while the antivortex-like structure consists of those aligned clockwise.
They are parameterized by the angle $0\leq \theta \leq \pi$ as
\begin{align}
    \bm{n}_A &=  \left(0,\sin\theta,\cos\theta \right), \\  
    \bm{n}_B &=  \left(-\frac{\sqrt{3}}{2}\sin\theta,-\frac{1}{2}\sin\theta,\cos\theta \right), \\ 
    \bm{n}_C &=  \left(\frac{\sqrt{3}}{2}\sin\theta,-\frac{1}{2}\sin\theta,\cos\theta \right),
\end{align}
for the vortex-like structure,
and
\begin{align}
    \bm{n}_A &=  \left(0,\sin\theta,\cos\theta \right), \\  
    \bm{n}_B &=  \left(\frac{\sqrt{3}}{2}\sin\theta,-\frac{1}{2}\sin\theta,\cos\theta \right), \\ 
    \bm{n}_C &=  \left(-\frac{\sqrt{3}}{2}\sin\theta,-\frac{1}{2}\sin\theta,\cos\theta \right),
\end{align}
for the antivortex-like structure.

With the above definition of magnetic textures,
we calculate the eigenenergies of the electrons under each possible magnetic texture,
and evaluate the total energy of the system.
Since $y$-direction in this system is symmetric under translations by unit cells for both types of magnetic textures,
the total Hamiltonian $H$ is diagonal in the momentum component $k_y$.
On the other hand,
translational symmetry in $x$-direction is violated by the spiral structures,
and hence we numerically diagonalize the lattice Hamiltonian to obtain the set of eigenenergies $\{ E_n(k_y) \}$.
From the obtained eigenenergies, we evaluate the total energy of the system,
\begin{align}
    U_{\mathrm{el}} &= \sum_{n,k_y \in \mathrm{occ.}} E_n(k_y),
\end{align}
where the sum is taken over the occupied states up to the given filling factor $\nu$.
By evaluating $U_{\mathrm{el}}$ for all the possible magnetic textures,
we determine the ground-state magnetic texture that minimizes $U_{\mathrm{el}}$,
for the given parameters $\nu$, $\lambda_{\mathrm{R}}$, and $\lambda_{\mathrm{KM}}$.

\section{Fitting with effective spin Hamiltonian}
The quantities $J_{1,2}$, $K_A$, and $D^{\parallel,\perp}$ are derived by fitting the total energy $U_{\mathrm{el}}$ calculated above to the effective spin Hamiltonian
\begin{align}
    H_S &= -\sum_i K_A (n_i^z)^2 \\
    & \quad + \sum_{\nn{ij}} \left[-J_1\bm{n}_i \cdot \bm{n}_j + \bm{D}^{\parallel}_{ij}\cdot (\bm{n}_i \times \bm{n}_j ) \right] \nonumber \\
    & \quad + \sum_{\nnn{ij}}\left[-J_2\bm{n}_i \cdot \bm{n}_j +\bm{D}^{\perp}_{ij}\cdot (\bm{n}_i \times \bm{n}_j ) \right]. \nonumber
\end{align}
$J_1$ and $J_2$ parametrizes the Heisenberg exchange interactions between nearest neighboring sites $\nn{ij}$ and between next-nearest neighboring sites $\nnn{ij}$.
$K_A$ is the parameter for the uniaxial anisotropy,
which prefers OOP orders if it is positive and IP orders if it is negative.

\begin{figure}[bp]
    \centering
        \includegraphics[width=1.0\linewidth]{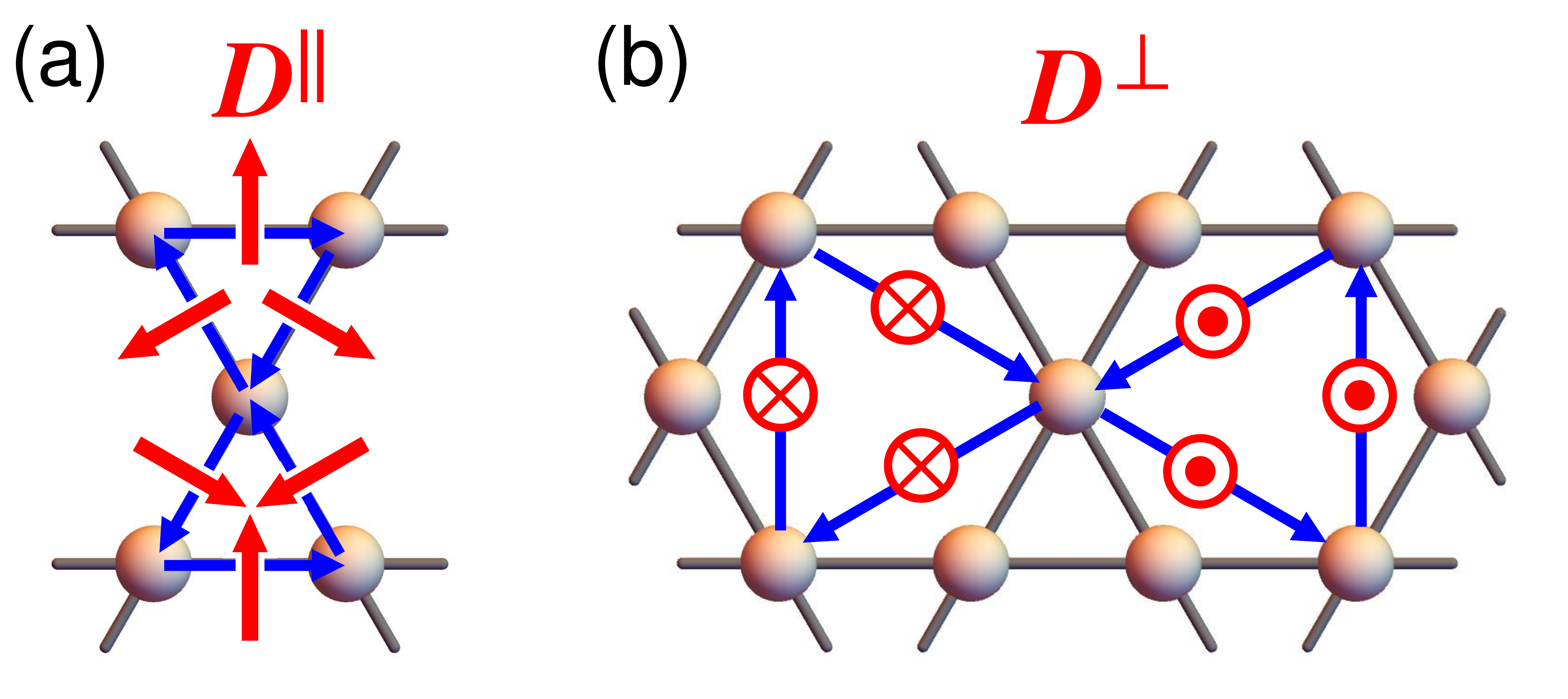}        
    \caption{
        (Color online)
        Schematic pictures of the DM interaction vectors taken in the effective spin Hamiltonian $H_S$: (a) the IP component $\bm{D}^{\parallel}_{ij}$ between nearest neighboring sites and (b) the OOP component $\bm{D}^{\perp}_{ij}$ between next-nearest neighboring sites.
        The red arrow on each link specifies the direction of the DM vector $\bm{D}^{\parallel,\perp}_{ij}$ for positive $D^{\parallel,\perp}$,
        where the sites $i$ and $j$ correspond to the head and tail of each blue arrow, respectively.
    }
    \label{fig:Fig_S3}
\end{figure}

The Dzyaloshinskii--Moriya (DM) interaction vectors $\bm{D}_{ij}^{\parallel,\perp}$ are introduced to quantify the tendency of reaching the noncollinear orders seen in the phase diagram.
Here we first fix their directions by the symmetry of the system,
and then determine their magnitudes by the fitting calculation.
The directions of $\bm{D}_{ij}^{\parallel,\perp}$ on each link are determined by the Moriya's rule as shown in Fig.~\ref{fig:Fig_S3}.
The IP component $\bm{D}_{ij}^{\parallel}$ originates from the global breaking of inversion symmetry in the OOP direction,
which is characterized by the Rashba SOC $\lambda_{\mathrm{R}}$ at surfaces or interfaces.
Its direction is thus defined as $\bm{D}_{ij}^{\parallel} \propto \hat{\bm{z}} \times \bm{e}_{ij}$.
On the other hand, the OOP component $\bm{D}_{ij}^{\perp}$ originates from the local breaking of inversion symmetry in the IP direction due to the lattice structure,
which is characterized by the KM-SOC $\lambda_{\mathrm{KM}}$.
Its direction is defined as $\bm{D}_{ij}^{\perp} \propto  \bm{e}_{kj} \times \bm{e}_{ik}$,
where $k$ denotes the nearest neighboring site between $i$ and $j$.

With the settings defined above,
we fit the calculated total energy $U_{\mathrm{el}}$ to the effective spin Hamiltonian $H_S$,
to determine the coefficients $J_{1,2}$, $K_A$, and $D^{\parallel,\perp}$ as functions of $\nu$ and $(\lambda_{\mathrm{R}},\lambda_{\mathrm{KM}})$ as shown in the main article.

\begin{figure}[tbp]
    \centering
        \includegraphics[width=1.0\linewidth]{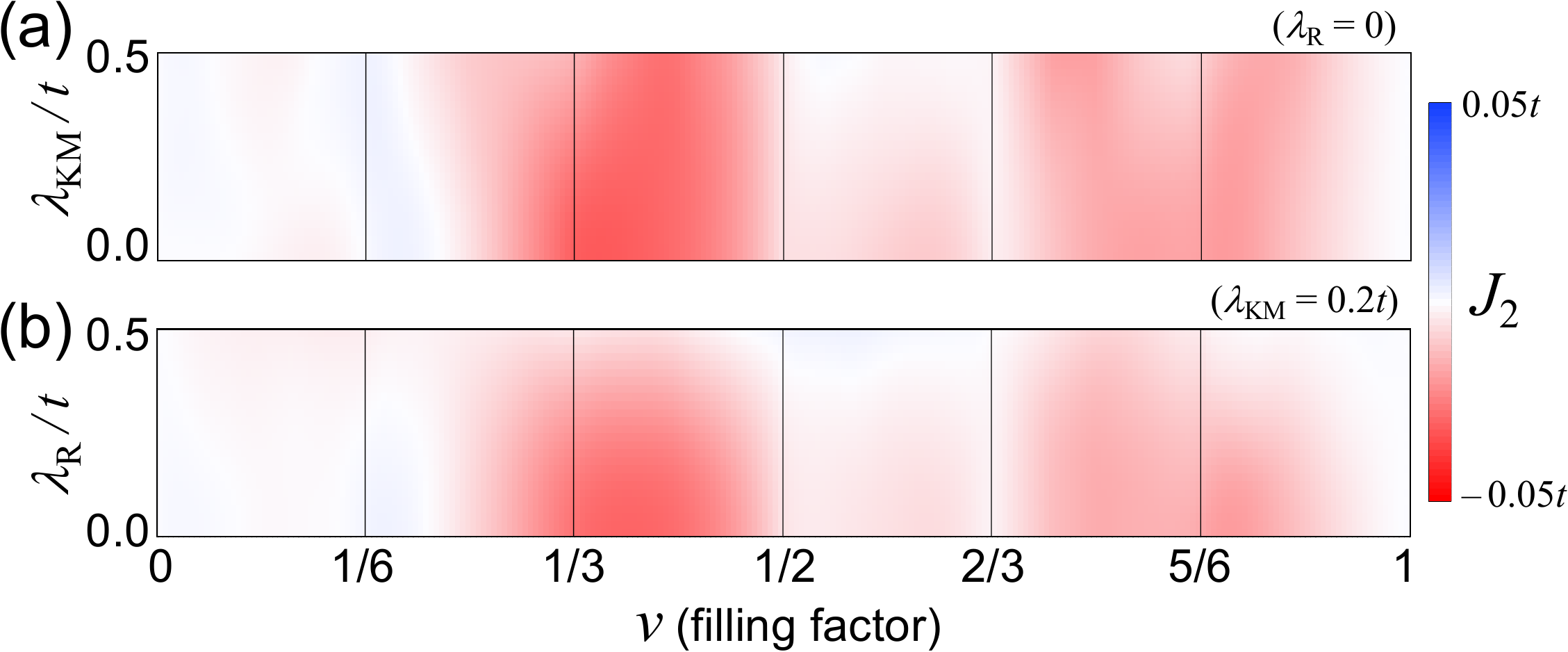}        
    \caption{(Color online)
        The numerically obtained behavior of the effective spin-spin interaction $J_2$ between next-nearest neighboring sites,
        (a) with $\lambda_{\mathrm{KM}}$ varied and $\lambda_{\mathrm{R}} =0$ fixed,
        and (b) with $\lambda_{\mathrm{R}}$ varied and $\lambda_{\mathrm{KM}} =0.2t$ fixed.
    }
    \label{fig:J2}
\end{figure}

Here we supplement the calculation results that are not shown in the main article.
The behavior of the effective coupling $J_2$ between next-nearest neighboring sites is obtained as shown in Fig.~\ref{fig:J2}.
We find that the coupling becomes weakly antiferromagnetic in almost all the parameter region.
Its strength is about one order of magnitude smaller than the nearest-neighbor coupling $J_1$.
Therefore, we can conclude that its effect on the ground-state magnetic texture is negligible,
except for the spiral state emerging between the FM and AFM states in the phase diagram.

\begin{figure}[tbp]
    \centering
        \includegraphics[width=1.0\linewidth]{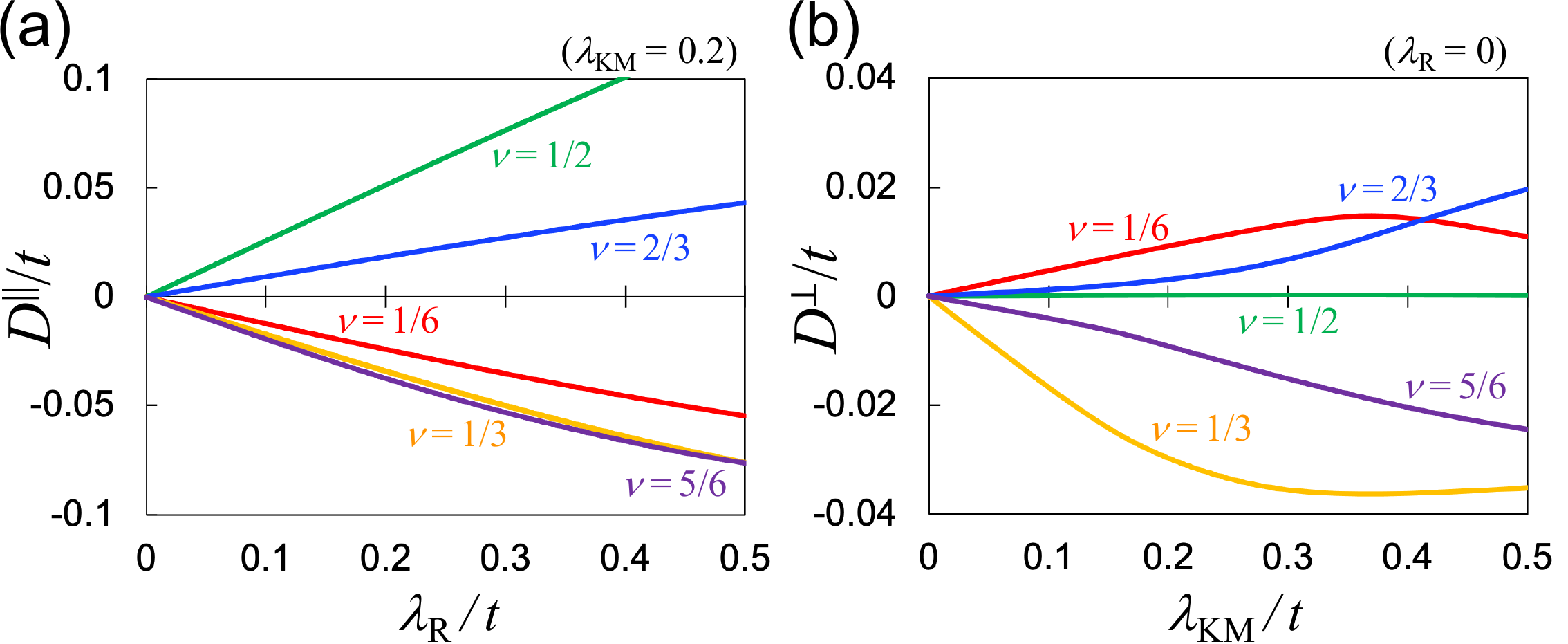}        
    \caption{(Color online)
        The numerical behaviors of the DM coefficients
        (a) $D^\parallel$ as a function of $\lambda_{\mathrm{R}}$
        and (b) $D^\perp$ as a function of $\lambda_{\mathrm{KM}}$,
        with the filling factor $\nu$ fixed.
    }
    \label{fig:DM}
\end{figure}

In order to discuss the origin of the effective DM interaction,
here we fix the filling factor $\nu$,
and plot the obtained DM coefficients $D^\parallel$ and $D^\perp$ as functions of the SOC coefficients $\lambda_{\mathrm{R}}$ and $\lambda_{\mathrm{KM}}$,
as shown in Fig.~\ref{fig:DM}.
As discussed in the main article,
the IP component $D^\parallel$ depends linearly on $\lambda_{\mathrm{R}}$ around $\lambda_{\mathrm{R}}=0$ in both the FM and AFM regimes.
On the other hand, we find that the OOP component $D^\perp$ depends linearly on $\lambda_{\mathrm{KM}}$ and $\lambda_{\mathrm{KM}}=0$.
We can thus conclude that $D^\parallel$ and $D^\perp$ arise as the consequences of the inversion symmetry breaking by $\lambda_{\mathrm{R}}$ and $\lambda_{\mathrm{KM}}$, respectively.
The signs of $D^\parallel$ and $D^\perp$ strongly depend on the filling factor $\nu$.



